# THE EXO-UV PROGRAM: LATEST ADVANCES OF EXPERIMENTAL STUDIES TO INVESTIGATE THE BIOLOGICAL IMPACT OF UV RADIATION ON EXOPLANETS

**Abrevaya, X. C.** [1,2,3,*]; **Odert, P.** [3]; **Leitzinger, M.** [3]; **Oppezzo, O.** [4,*]; **Luna, G.J.M.** [5,*]; **Patel, M. R** [6]; **Hanslmeier, A.** [3]

[1]*Instituto de Astronomía y Física del Espacio (UBA - CONICET), Pabellón IAFE, Ciudad Universitaria, C1428EGA, Ciudad Autónoma de Buenos Aires, Argentina.*
[2] *Facultad de Ciencias Exactas y Naturales, Universidad de Buenos Aires, Ciudad Universitaria, C1428EGA, Ciudad Autónoma de Buenos Aires, Argentina.*
[3]*Institute of Physics, Department of Astrophysics & Geophysics, University of Graz, Universitätsplatz 5, 8010, Graz, Austria.*
[4]*Departamento de Radiobiología, Comisión Nacional de Energía Atómica, Centro Atómico Constituyentes, B1650KNA, Buenos Aires, Argentina.*
[5]*Universidad Nacional de Hurlingham (UNAHUR), Secretaría de Investigación, Av. Gdor. Vergara 2222, B1688GEZ, Villa Tesei, Buenos Aires, Argentina.*
[6]*Consejo Nacional de Investigaciones Científicas y Técnicas (CONICET), Godoy Cruz 2290, C1425FQB, Ciudad Autónoma de Buenos Aires, Argentina.*
[7]*School of Physical Sciences, The Open University, Milton Keynes, MK7 6AA, UK.*
*Members of the Argentinian Research Unit in Astrobiology

**Abstract**

The EXO-UV program is an international, interdisciplinary collaboration between astrophysicists and biologists aimed at expanding the characterization of ultraviolet radiation (UVR) environments on exoplanets. This approach combines astrophysical studies with biological experiments to better understand the potential impacts of UVR on exoplanetary surfaces. UVR is particularly relevant because it reaches the surface of planets and can influence their habitability. The specific wavelengths within the UVR spectrum depend on the planet's atmospheric composition and the spectral energy distribution of its host star. Additionally, high UVR fluxes emitted during flares and superflares are of particular interest due to the limited information available regarding their biological impact. The EXO-UV program has successfully led to the first experimental study examining the biological effects of high UVR fluences, such as those produced by flares and superflares. This initial study focused on Proxima b, the closest potentially habitable exoplanet to our solar system, which orbits the M-dwarf star Proxima Centauri. Building on this, our next focus within the program was the TRAPPIST-1 system, which has captured significant scientific attention since its discovery in 2017. This system, orbiting the ultra-cool dwarf star TRAPPIST-1, contains seven planets, three of which are considered to be within the habitable zone (planets e, f, and g). Many studies have explored the potential for life on these planets, particularly investigating the impact of stellar UVR on their surfaces, both during quiescent and flare flux conditions. Previous studies often relied on calculations derived from existing data on microorganisms irradiated at low fluence rates and fluences. In contrast, our approach involved conducting specific laboratory experiments to re-evaluate the findings from earlier studies. Our results suggest that previous research underestimated the ability of "life as we know it" to withstand these fluences, whether for UVR-tolerant or UVR-susceptible microorganisms. Future experimental studies aim to investigate the biological effects of repetitive flares. In this paper, we review the latest results from our EXO-UV program.

1. Introduction

In the past decades, technological advancements have led to an exponential increase in the detection of exoplanets, expanding the boundaries of space exploration within our galaxy. As of today, thousands of exoplanets have been discovered, including planetary bodies with masses and radii consistent with rocky planets orbiting nearby low-mass stars (e.g., Anglada-Escudé et al., 2016; Quintana et al., 2014; Borucki et al., 2013, among others).

Simultaneously, there has been a growing interest in assessing the potential habitability of these planetary bodies—specifically, the characterization of environments beyond Earth that may be capable of supporting life (Lammer et al., 2024; Scherf et al., 2024). The research about habitability has become one of the primary goals of major international astrobiology research programs (e.g., Des Marais et al., 2008; Horneck et al., 2016).

Exoplanets considered as potential hosts for life are typically those cataloged as "Earth-like" and are located within the classical habitable zone—the region around a star where liquid water can exist on a planet's surface (Kasting et al., 1993; Kopparapu et al., 2013). These planets may orbit various types of stars, particularly F, G, K, and M dwarf stars, whose main-sequence lifetimes are long enough to allow for the development of life. The abundance of these stars varies across our galaxy, with F, G, and K stars accounting for approximately 20% of the stellar population, while M dwarfs make up around 75% (Jiang et al., 2024).

Although G stars are relevant targets in the search for life due to their similarities with the Sun, which could lead to conditions resembling those present on Earth, the search for exoplanets is shifted to M stars as those are approximately eight times more abundant, meaning that a significant fraction of the discovered exoplanets orbit these stars. Apart from their abundance, it is much easier to detect planets orbiting M stars because of their lower luminosities. Statistically, each M-star has 1.3 ± 0.3 planets with 1 to 10 MEarth (Sabotta et al. 2021). However, according to recent studies, it may be easier to detect habitable planets on K-type stars rather than M-type stars or G-type stars. Searching for habitable planets orbiting K-type stars implies a compromise between technical and physical feasibilities (Lillo-Box et al., 2022). Planets in the HZ of G-type stars have periods of around 1 year, thus making detection difficult. The period for planets in the HZ around K-type stars is shorter than for G-type stars. However, K-type stars are not as active as M-type stars and their HZ planets are not endangered to be tidally locked. Again, this displaces the interest in studying M-type stars, apart from the facts mentioned above. Interestingly, the environment of planetary bodies orbiting M-dwarfs can differ significantly from that of Earth. Because M-dwarfs are cooler than the Sun, their habitable zones are located much closer to the star. If the host star is more active than the present-day Sun this could increase the high-energy radiation impact on the planet (Scalo et al., 2007; Dressing & Charbonneau, 2015). This increased high-energy radiation could ultimately influence the planet's environment and shape the potential for life to develop.

Essential requisites for planetary habitability (as a measure of the potential of a planetary body to develop and maintain an environment hospitable to life for a significant period of time) are not only limited to the presence of liquid water, organic molecules, and energy sources fundamental for life "as we know it". It also involves the evaluation of key factors that are known to be critical for life forms considered as potential inhabitants (Cockell, 2007). These include the evaluation of the impact of stellar radiation on the emergence and survival of life on planetary bodies which is fundamental for determining the potential of a planetary body to harbor life (Khodachenko et al., 2007; Lammer et al., 2007).

Among the factors constraining planetary habitability, the evaluation of stellar radiation particularly UV wavelengths is fundamental because it can determine the feasibility of a planet to harbor life on its surface. This is because UVR (200-400 nm) can reach the surface of a planet having direct or indirect effects on life (Abrevaya & Thomas, 2018). In general, potential life forms on other planets are considered simple life forms, such as microorganisms. As direct effects, we can consider those that UVR can have on life forms (or prebiotic molecules). These effects could be beneficial but others could be detrimental for life (e.g.: Kielbassa et al., 1997). As indirect effects, we can consider those that can affect the planetary environment, in particular the atmosphere. It is well known that the radiation and plasma environments of the host star play a role in the evolution of a planet and its atmosphere (e.g. Khodachenko et al., 2007). Moreover, the planetary atmospheric composition and pressure alters the fluxes of stellar radiation reaching the planetary surface, which can influence the impact on life. The stellar UVR depends on the stellar spectral type and magnetic activity. Stellar activity manifests itself as energetic events such as flares and coronal mass ejections (CMEs), as well as enhanced coronal X-ray and chromospheric ultraviolet (UV) emission (e.g. Ayres, 1997; Gershberg, 2005; Scalo et al., 2007). Particularly flares and superflares (flares with an energy $>10^{33}$erg) are sudden energetic radiation outbursts originating in magnetic reconnection processes having responses in all layers of the stellar atmosphere (photosphere, chromosphere, and corona).

The study of UV habitable environments on extrasolar planets is a highly active area of research (for reviews on the topic, see e.g.: Hanslmeier, 2018; Abrevaya & Thomas, 2018). However, few studies have assessed the biological impact of (super) flares on planetary bodies. Ranjan et al. (2017) highlighted the lack of experimental approaches to studying UV habitability on extrasolar planets and the need for the development of more experimental studies under laboratory conditions. Moreover, previous research has largely relied on theoretical estimations of these biological effects, often based on calculations derived from existing experimental data. To date, information on the biological effects of radiation has primarily been obtained from studies conducted at low fluences and low fluence rates in the UV range and has not directly addressed the impact of (super)flares on planetary bodies using specific experiments designed to this end.

We can define theoretical approaches that involve the use of experimental data from the literature, along with mathematical models, computer simulations, and other methods to estimate biological effects. On the other hand, the experimental approach involves exposing terrestrial microorganisms (or molecules) to extraterrestrial conditions. Within this approach, there are two main types of experiments: space experiments, conducted in real environmental conditions to simulate various scenarios, typically performed in Low Earth Orbit (LEO), and laboratory experiments, conducted under simulated environmental conditions using specialized facilities and instruments.

As part of the EXO-UV program (Abrevaya et al., 2014), we propose a combined approach that integrates astrophysics and biology through theoretical and observational astrophysical studies combined with simulation experiments under laboratory conditions to assess the biological impact of radiation. This experimental approach involves the use of simulation chambers, and facilities such as accelerators or irradiation devices (many of which are custom-built, with few being commercially available).

Therefore the EXO-UV program is an interdisciplinary project that integrates biology and astrophysics aiming to study the effects of stellar UVR from F, G, K, and M stars on potential life forms on exoplanets to expand and improve previous studies by incorporating experimental laboratory approaches.

**2. Biological impact of UVR in the astrophysical context**

As mentioned earlier, UVR is a key factor in habitability studies as it can reach a planet's surface (the wavelengths reaching the surface will depend on the surface pressure and composition of an atmosphere). UVB and UVC radiation can be particularly harmful to organisms (e.g.: Kielbasa 1997). Although the biological effects of UVR have been studied for decades, much of this research has focused on terrestrial scenarios or sterilization processes in Earth-like environments, which typically involve low fluence rates and fluences.

While it can be argued that life may exist beneath the surface in highly irradiated environments and therefore shielded from high levels of ultraviolet radiation (UVR), the search for exoplanet biosignatures is at least partially focused on the possibility of having life on the surface (surface biosignatures, see e.g.: Hedge et al., 2015). Therefore, this emphasizes as well the need for studies to assess the surface habitability of exoplanets and find out whether UVR could pose a threat to life on the surface of the planet.

Even though several studies have investigated stellar UVR to evaluate the potential habitability of exoplanet surfaces, most rely on data taken from the literature (e.g., Cockell 1999; Buccino et al. 2006, 2007; Fossati et al. 2012; O'Malley-James & Kaltenegger, 2017; Howard et al. 2018; Estrela & Valio 2018; Rugheimer et al. 2015; O'Malley-James & Kaltenegger, 2019). These studies typically use biological action spectra (BAS) and extrapolate values for microbial survival from parameters such as the lethal dose (LD) or the fluence required to reduce a percentage of the microbial population (F). BAS describes the relative effectiveness of radiation at different wavelengths in producing a biological effect, such as damage. However, they do not directly indicate the survival chances of a microbial population, but instead provide information on the wavelengths at which radiation is most damaging. BAS has been used to estimate the UV surface habitability of exoplanets, focusing on microorganisms (Rugheimer et al., 2015; O'Malley-James & Kaltenegger, 2017; Estrela & Valio, 2018; O'Malley-James & Kaltenegger, 2019; Estrela et al., 2020) or DNA molecules in vitro (Buccino et al., 2006, 2007; Rugheimer et al., 2015). The LD represents the amount of radiation (dose) needed to kill a certain percentage of a microbial population (e.g., $LD_{90}$, indicates the amount of radiation required to kill 90% of the population), meanwhile, F represents the amount of radiation (fluence) to obtain a certain percentage of survival (e.g., $F_{10}$, indicates the amount of radiation required to obtain 10% of survival). These values are usually obtained from the exponential part of the survival curves (see Fig. 1). The use of these values assumes that the lethal effect is an exponential function of the imparted dose for a given microorganism. The LD and F values have been used to estimate the impact of UVR on life on exoplanets during superflares (Howard et al., 2018). Although UV-BAS and LD (or F) provide insights into the biological effectiveness of UVR and microbial survival, respectively, they have limitations for the evaluation of the UVR effects on life, especially if they are used to predict survival in a higher range of fluences than those involved in the experiments from where these values are obtained.

Many models fail to account for critical factors, leading to (i) inaccurate estimates of UV flux at the planet's surface, often neglecting the presence of a planetary atmosphere or assuming unrealistic atmospheric compositions for the stellar spectral type and activity (for example, $O_2$-rich Earth-like atmospheres are unlikely on planets orbiting M dwarfs, as shown by Johnstone et al. 2019); and (ii) incorrect estimations of the biological impact of UVR, often relying on extrapolations from UV-BAS or UV-LD data, which measure biological damage at specific wavelengths under constant fluence or estimate the radiation needed to kill a population, without fully considering the diverse environmental conditions of exoplanets.

The main limitations of these approaches can be classified as astrophysical and biological and can be described as follows:

-Astrophysical limitations:

Some of these studies did not assume either the presence of a planetary atmosphere or atmospheric compositions that are unlikely to exist on planets orbiting stars of a certain stellar spectral type and activity level (e.g.: $O_2$ rich Earth-like atmospheric compositions are likely unfeasible in planets orbiting dM stars, e.g.: Garcia-Sage et al., 2017; Johnstone et al., 2019). Moreover, they were mostly based on synthetic stellar spectra, not on actual observations. In this context, both flare activity and flare frequency have been very poorly studied. In the UV, there is more limited data compared to the optical range, especially observations of flares and superflares, because observations need to be carried out from space. The bluest photometric band reachable from the ground covers the biologically relevant UVA band (300-400nm), but UVB and UVC are only accessible via satellite observations. However, detailed UV data are relevant for photochemical processes in planetary atmospheres which affect the atmospheric conditions under such irradiation levels.

The interaction of planets with flare emissions, which can be up to several orders of magnitude higher compared to the quiescent state of a star, is not well understood and it remains unclear whether it could be unfavorable for life. These high levels of UV radiation might be lethal for life forms, especially for close-in orbiting planets which may have lost a large amount of their "life-protecting" atmospheres. This is particularly important for dM stars and their high-flaring activity.

-Biological limitations:

1) The values used in the extrapolations made in previous research to investigate the UVR habitability of exoplanets are taken in general from studies that aim to evaluate the action of sunlight on terrestrial ecosystems (where, for instance, the harmful effects of UVC radiation are not present), or from studies focusing on the effect of UVR on microorganisms using sunlight or artificial UV sources for disinfection processes or radio-sensitivity studies (where the UV fluence rates and fluences are in general much lower than those from flares/superflares, see e.g.: Gascón et al., 1995; Nelson et al., 2018; Szeto et al., 2020). Because the characteristics of the incident radiation (wavelength, fluence rate, and fluence) will determine the biological effects the possible biological impact of UV during stellar flare events cannot be extrapolated from these data in the literature.

2) Another factor, usually not considered, is that the survival of the microorganisms varies according to their growth stage (e.g. growth phase) and the physiological conditions of the population (e.g. starvation). Therefore extrapolations from experiments mentioned in (1) are difficult as they were performed with a set of particular conditions.

3) Approaches considering damage to DNA molecules (as DNA-BAS) can be unrealistic, as DNA is not the only component that can be affected by radiation damage, proteins and other cellular structures can also be damaged. This approximation also does not take into account the cellular DNA repair mechanisms and other intrinsic mechanisms of the cell to cope with UVR, such as photoprotection. In general, the probability of surviving UVR exposure does not only depend on the effectiveness of the damage, but on the balance between photoprotection, damage, and DNA repair (e.g. Jones & Baxter, 2017). These mechanisms are particularly important in microorganisms that are resistant or tolerant to UVR, known as radioresistant or radiotolerant, respectively.

4) When using the UV-LD the extrapolations are made from the survival curves, the shape of the curve will be influenced by the experimental factors mentioned in (1) and (2). At high UV fluences, like those expected during stellar flares or superflares, survival curves are frequently biphasic or multiphasic (Fig. 2, right panel). Therefore the effect of UVR on life cannot be appraised from both, the initial slope of the curve and the total fluence imparted, such as previous studies have assumed which used the UV-LD.

Through the first results obtained in the EXO-UV program (Abrevaya et al. 2020; Abrevaya et al., 2024) we determined that previous studies based on these approaches underestimated the chances of life as we know it to survive the high UV fluences of flares and superflares.

Therefore, new studies based on interdisciplinary experimental approaches are needed to more realistically determine the chances of life to arise or develop on a planet, as well as to estimate biologically relevant UV fluences.

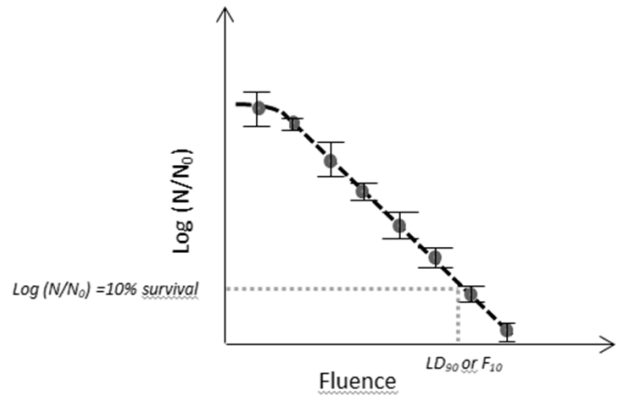

Figure 1. Typical survival curve or microorganisms exposed to UVR obtained at low fluences exhibiting an exponential decay. Values such as LD90 or F10 are calculated from the exponential part of the curve considering the 10% survival (this curve is shown only for illustrative purposes and does not contain experimental data).

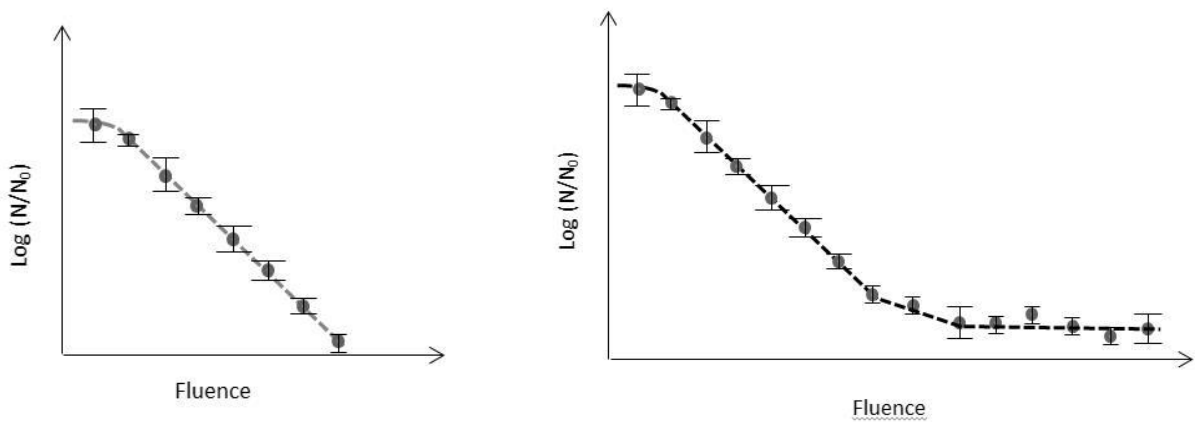

Figure 2. Comparison of the typical survival curves obtained for microorganisms exposed to UVR at low and high fluences. The left panel shows a survival curve for low fluences (lower than those typically encountered during a flare event), showing a simple exponential decay. The right panel shows a survival curve for high fluences (similar to those experienced during a flare.), showing a multiphasic curve (these curves are only shown for illustrative purposes and do not contain experimental data).

### 3. General methodology of the EXO-UV program

As part of the EXO-UV program, we use a combined methodology, which involves an astrophysical analysis and biological experimental studies.

For the astrophysical part, we apply an astrophysical analysis which provides the stellar UV fluxes to be used in the biological laboratory experiments. This includes gathering data on UV quiescent levels, flares, and superflares of representative dM stars. At the first stage of the program, we selected systems with already known rocky planets (Proxima Centauri, Trappist-1) and searched for existing data covering the UV range for these stars as well as used extrapolations from the literature. Possible sources of UV data for the EXO-UV program include archives from the Hubble Space Telescope (HST) and the International Ultraviolet Explorer (IUE), which provide spectroscopic observations covering (partly) the biologically relevant UV range (200-400 nm). For several dM-type planet host stars, full spectral energy distributions from X-rays to the infrared are readily available from the MUSCLES program (France et al., 2016; Wilson et al. 2024). Furthermore, photometric data from GALEX can be used to constrain the UV emission from dM stars (Shkolnik & Barman, 2014), whose NUV filter band (~177-283 nm) includes UVC.

For the case of flares/superflares, observations in the UV are not that common, but observations of individual flares on dM stars do exist, e.g. from IUE (Hawley & Pettersen, 1991) or GALEX (Welsh et al., 2007). Recently, several studies searched the GALEX data for flares (e.g. Million et al. 2023, Rekhi et al. 2023, Berger et al. 2024) and provided near-UV flare statistics.

An alternative approach is to use the more abundant optical flare frequency distributions (FFDs) of dM stars from the literature (e.g. from Kepler and TESS; Hawley et al., 2014; Günther et al., 2020). The drawback of using Kepler or TESS FFDs is that their photometric bands do not cover the UV range. Therefore, it will be necessary to extrapolate the corresponding UV emissions of these flares. The optical/UV continuum emission in flares can be approximated by a blackbody with a temperature around 10000 K rather well (e.g. Hawley et al., 2003; Howard et al., 2018). However, recent studies revealed that the UV fluxes may be underestimated by factors of a few with this simple assumption (Jackman et al. 2023, Brasseur et al. 2023).

The UV fluxes in quiescence, as well as during flares and superflares, are scaled to the habitable zone around these stars, which are calculated following Kopparapu et al. (2013), leading to a range of unshielded UV flux values and energies.

In principle, we consider the worst-case scenario without any source of shielding, i.e: microorganisms would receive unshielded UV radiation, but depending on the stellar UV flux levels potential planetary atmospheric compositions and radiative transfer modeling to evaluate surface fluxes with atmospheric shielding can be considered. A 1-D radiative transfer modeling can be used to obtain the UV fluxes at the surface of the planetary bodies for different atmospheric compositions and pressures (e.g. Patel et al. 2002). Atmospheric compositions of Earth-like exoplanets orbiting in the habitable zone of dM stars are not yet constrained by observations. Therefore, different hypothetical atmospheres for such planets can be considered, focusing on Earth-like planets with different atmospheric compositions and pressures. For instance, mixtures of $CO_2$ and $N_2$ atmospheres are likely feasible for exoplanets orbiting dM stars according to photochemical studies in the literature (Rugheimer et al., 2015; Ranjan et al., 2017).

For the biological experiments we used a standard bench scale apparatus, consisting of low-pressure Hg lamps enclosed in pseudo-collimated irradiation systems (Quals & Johnson 1983; Bolton & Linden, 2003), but the fluence rates and exposure times were set to values considerably higher than those usually assayed with this kind of device, resembling the conditions expected to occur in exoplanets during flares and superflares. For the estimation of the required fluences, we considered the UVC (200-280 nm) and UVB (280-315 nm) components of the incoming radiation since these bands are the most damaging for microorganisms, and we can assume that most of the damage produced by UV will come from these wavelengths. The UV output emission was measured with a standard radiometer or using iodide/iodate actinometry (Rahn 1997) when the samples were close to the source of radiation in order to increase the fluence rate. In the latter case, the samples are placed on a support cooled by water circulation to avoid temperature increases. The survival is evaluated by examining the ability of microorganisms to form colonies in solid medium following exposure and quantified by the number of colony-forming units (CFUs). The survival fractions are obtained by counting the number of CFUs of unirradiated samples (control group, $N_0$) and the number of CFUs of irradiated samples (N), and calculated as $N/N_0$. The survival curves as obtained plotting the survival fraction as a function of the applied fluence.

### 4. First results of the EXO-UV program

**4.1 Proxima b**

In this program, we took as a starting case study the planet Proxima Centauri b (Proxima b) (Abrevaya et al., 2020), the closest potentially habitable terrestrial-mass exoplanet outside the Solar System. Situated within the habitable zone of Proxima Centauri, an M6V dwarf star (Anglada-Escudé et al., 2016), Proxima b represents an intriguing target for exploration.

Previous studies that intended to disentangle the UV habitability of this planet based on models predicted that microbial life could exist only if protected under an ozone atmosphere (O´Malley James & Kaltenegger, 2017) and predicted no survival to superflares, taking as a parameter for lethality fluences greater than 553 $Jm^{-2}$ (e.g.: Howard et al., 2018).

We studied both quiescent and flaring stellar conditions. For our analysis of quiescent conditions, we applied radiative transfer modeling to estimate the UVR at the surface of Proxima b, taking into account different atmospheric compositions and pressures. We assumed planetary atmospheres composed of $CO_2$ and $N_2$, with surface pressures ranging from 100 to 5000 mbar, based on several studies (Tian et al. 2008, Lichtenegger et al. 2010, Lammer et al. 2011, Gao et al. 2015, Johnstone et al. 2018; 2019). Our results have shown that these atmospheric compositions and pressures can provide shielding from the most harmful UVR wavelengths, thereby broadening the range of "UV-protective" atmospheric compositions beyond solely ozone. This shielding increases as a function of the $CO_2$ abundance and as a function of the pressure.

Furthermore, our results show that the UVR reaching the surface of Proxima b under quiescent conditions would be negligible from a biological perspective, even in the absence of an atmosphere (Abrevaya et al., 2020).

We then decided to investigate the effects of flaring conditions, since high UV fluxes could pose a significant threat to the potential existence of life, we experimentally examined the impact of flares on microorganisms under a "worst-case scenario" (i.e., no UV shielding) to assess the biological damage caused by flares on the surface of Proxima b, particularly within the UVC range, the most biologically harmful type of radiation. These experiments included microorganisms from different domains of life: *Haloferax volcanii*, an archaeon that thrives in extreme hypersaline environments, and *Pseudomonas aeruginosa*, a ubiquitous bacterium. For the irradiation procedure, suspensions of these microorganisms were exposed to UVC tube lamps at fluence rates of 8.7 W m$^{-2}$ (typical flare) and 92 W m$^{-2}$ (superflare).

Our results demonstrate the impact that both typical flares and superflares could have on life, in particular, microorganisms. When exposed to extremely high UVC fluences—such as those expected to reach the surface of Proxima b during a typical flare or superflare—a small fraction of the microbial population can survive, regardless of the species (for comparison and summary of the results, see Table 1). Notably, species like *H. volcanii* and *P. aeruginosa*, which are generally considered less radiotolerant than *D. radiodurans* (a standard for radiotolerance in previous studies), have shown in this experiment that they may still have chances to survive these events.

|  | Proxima b | |
|---|---|---|
|  | **Howard et al., 2018** | **Abrevaya et al., 2020** |
|  | *Unshielded* | *Unshielded* |
| Different microorganisms | <u>Superflare</u><br><br>*D. radiodurans* survival constrained to LD90 553 Jm$^{-2}$<br><br>More complex life forms such as Lichens would have to undergo special adaptations to survive (even with protection of planetary atmosphere) | <u>Flare</u><br><br>*H. volcanii* survival up to the LOD, after exposure to an equivalent fluence of a flare ($10^{-6}$)<br><br>*P. aeuruginosa* survival after exposure to an equivalent fluence of a flare ($10^{-6}$-$10^{-7}$)<br><br><u>Superflare</u><br><br>*P. aeuruginosa* survival after exposure to an equivalent fluence of a flare ($10^{-4}$-$10^{-5}$) |

Table 1. Comparative results obtained for different estimations of the survival of different microorganisms on the surface of Proxima b, after being exposed to UVR fluences equivalent to a flare (Abrevaya et al., 2020) and a superflare (Howard et al., 2018 and Abrevaya et al., 2020). The results shown are related to the use of a theoretical estimation that employs the LD$_{90}$ values (Howard et al., 2018) and an experimental approach employed in the EXO-UV program (Abrevaya et al., 2020). For the latter, the survival fractions on each planet for each microorganism are shown between brackets.

**4.2 TRAPPIST-1 e, f and g**

The second case study of our program focused on the TRAPPIST-1 system (Abrevaya et al., 2024), composed of seven Earth-like planets (b, c, d, e, f, g, and h) orbiting a nearby ultracool M8 dwarf star, TRAPPIST-1, which is located 12 pc away from Earth (Gillon et al. 2017). Planets e, f, and g are within the liquid water habitable zone (LW-HZ), while planets b, c, and d may only have limited liquid water. Atmospheric studies of these planets have not revealed absorption signatures, and hydrogen-rich extended atmospheres have been ruled out. However, some models suggest that the two outermost planets could retain atmospheres over long timescales. Despite challenges such as stellar contamination, early observations with JWST suggest the presence of either bare rock surfaces or thin oxygen-dominated atmospheres on planets e, f, and g.

TRAPPIST-1 itself is an active, old dM star with a mass close to the hydrogen-burning limit. It exhibits flare activity, with a flare rate of one every two days, and some flares have superflare energy levels (Vida et al. 2017). Studies of the system's habitability indicate that planet e is the most suitable for supporting life, with the possibility of liquid water oceans (Wolf, 2017; Dobos et al., 2019). Models also suggest that tidal heating could prevent planets d and e from entering a runaway greenhouse phase, while planets f, g, and h might have icy surfaces with subsurface liquid water (Dobos et al., 2019).

Recent research has focused on the biological impact of UVR from TRAPPIST-1 flares. Studies used microorganisms like *Deinococcus radiodurans* and *Escherichia coli*, analyzing parameters such as the $F_{10}$ to assess microbial survival after flare events. However, these studies often used data obtained at UV fluences lower than those expected during superflares (e.g.: Estrela et al., 2020).

In our study, we first calculated the UV fluxes at the surfaces of planets e, f, and g in the LW-HZ during both quiescent conditions and a superflare event. We determined that during quiescent conditions the UVR fluxes would be negligible. Then, in contrast with previous approaches that relied on theoretical models, we conducted experiments to assess the biological effects of high fluences of UVR on planets e, f, and g, considering a superflare case. The UV-tolerant bacterium *Deinococcus radiodurans* and the UV-susceptible bacterium *Escherichia coli* were exposed to fluences of germicidal radiation (254 nm) expected to produce a lethal effect equivalent to that of the radiation emitted during a superflare reaching the unshielded surfaces of these planets.

In contrast to the conclusions of previous studies, which estimated the microbial survival from $F_{10}$ values taken from the literature (Estrela et al., 2020), the results of our biological experiments demonstrated that a fraction of the microbial population could survive the intense UVR fluences of a superflare on the surface of TRAPPIST-1 planets, even in the absence of atmospheric or oceanic shielding and regardless to tolerance features of these microorganisms to UVR (for comparison and compilation of the results see Table 2).

|  | TRAPPIST-1 e, f, g | | | |
| --- | --- | --- | --- | --- |
|  | Estrela et al., 2020 | | | Abrevaya et al., 2024 |
|  | Conditions | | | Conditions |
|  | *Anoxygenic* | *Oxygenic (ozone)* | *Ocean* | *Unshielded* |
| ***E. coli*** | No survival | Survival | Survival only at 8, 9 and 11 m below the ocean for planets g, f, and e, respectively. | Survival on planet e, below the LOD<br>Survival on planet f ($10^{-9}$)<br>Survival on planet g ($10^{-9}$) |
| ***D. radiodurans*** | No survival on planet e Survival on f and g | Survival | Survival | Survival on planet e ($10^{-6}$)<br>Survival on planet f ($10^{-7}$)<br>Survival on planet g ($10^{-8}$) |

Table 2. Comparative results obtained for different estimations of the survival for a radiotolerant (*D. radiodurans*) and a susceptible microorganism (*E. coli*) on the surface of the TRAPPIST-1 planets e, f, and g, after being exposed to UVR fluences equivalent to a superflare (calculated by Vida et al., 2018). The results shown are related to the use of a theoretical estimation that employs the $F_{10}$ values (Estrela et al., 2020) considering different kinds of planetary environments, and an experimental approach employed in the EXO-UV program (Abrevaya et al., 2024). For the latter, the survival fractions on each planet for each microorganism are shown between brackets.

## 5. Discussion and conclusions

The results described in our program document the tolerance of well-known microorganisms to high fluences of UVR (in quantities and wavelengths that these microorganisms do not experience on present Earth). This can be correlated with the following factors. A microbial population, even a pure culture obtained from an isolated colony, is composed of cells with different characteristics. In populations of microorganisms exposed to UVR, the presence of small sub-populations of cells able to tolerate high fluences of UVR is usual and has been ascribed to aggregation (Kowalski et al. 2019) or transitory and reversible phenotypic changes (Pennell et al. 2008). When a microbial population is exposed to low fluences of UVR, most of the decrease in the number of cultivable cells results from the rapid inactivation of the susceptible cells, which are the majority of the overall population. This reduction of the survival fraction with increasing fluences usually resembles a simple exponential decay (see Fig. 2, left panel), and this is the kind of response that is assumed when values for $LD_{90}$ or $F_{10}$ are calculated. When a microbial population is exposed to high fluences, resembling those that they would receive from a flare, the existence of a small fraction of cells can tolerate these fluences (which is visible as a tail in the survival curves, see Fig. 2 right panel), meanwhile, the overall population seems to be inactivated. This minority sub-populations of tolerant cells are frequently disregarded (Cerf 1977), but due to their presence, the survival fractions at high fluences could be considerably higher than those predicted by extrapolation of the exponential decay observed at low fluences. The extrapolation of results from low fluences to high fluences could, therefore, lead to underestimating the ability of a microbial population to survive UVR exposures, and experimental determinations -under conditions resembling those expected in the environment under study- are required to evaluate this ability. Therefore, current findings from the EXO-UV program suggest that life could potentially endure highly UV-irradiated environments on exoplanets under conditions not found on present Earth. Part of the future aims of the EXO-UV program point towards an experimental demonstration of the ability of the minority population of survivors to generate a new microbial population comparable to the initial one, which is fundamental in evaluating whether extinction could be avoided. Moreover, as part of this another factor aimed to be analyzed in future experiments in this program involves evaluating microbial survival after exposure to repetitive flares and superflares.